\documentclass[aps,prb,twocolumn]{revtex4-1}  % for review and submission
\usepackage[version=3]{mhchem}	%chemical formulas
\usepackage{graphicx}  % needed for figures
\usepackage{nicefrac}  
\usepackage{siunitx} 	%Newer, better than SIunits, with ranges, numbers, units, etc.
\usepackage{amsmath,amssymb,amsfonts}   % for math
\usepackage{hyperref}
\usepackage{import}
\usepackage{mathdesign}
\usepackage{bbold}
\usepackage{wasysym}

\hypersetup{colorlinks=true,urlcolor=blue,citecolor=blue,linkcolor=blue,breaklinks}
\renewcommand{\vec}[1]{\mathbf{#1}}

\begin{document}

\title{Breaking Symmetry with Light: Ultra-Fast Ferroelectricity and Magnetism from Three-Phonon Coupling}

\author{Paolo G. Radaelli}
\affiliation{Clarendon Laboratory, Department of Physics, University of Oxford, Oxford, OX1 3PU, United Kingdom}
\email[Corresponding author: ]{p.g.radaelli@physics.ox.ac.uk}

\date{\today}

\begin{abstract}
A theory describing how ferroic properties can emerge transiently in the ultra-fast regime by breaking symmetry with light through three-phonon coupling is presented.  Particular emphasis is placed on the special case when two exactly degenerate mid-infra-red or THz phonons are resonantly pumped, since this situation can give rise to an exactly rectified ferroic response with damping envelopes of $\sim$ 1 ps or less.  Light-induced ferroelectricity and ferromagnetism are discussed in this context, and a number of candidate materials that could display these phenomena are proposed.  The same analysis is also applied to the interpretation of previous femto-magnetism experiments, performed in different frequency ranges (visible and near-infrared), but sharing similar symmetry characteristics. 
\end{abstract}

%\pacs{}
\maketitle

\section{Introduction}

The use of light to control the structural, electronic and magnetic properties of solids is emerging as one of the most exciting areas of condensed matter physics.  One promising field of research, known as femto-magnetism, has evolved from the early demonstration that magnetic `bits' in certain materials can be `written' at ultra-fast speeds with light in the visible or IR range. \cite{Bigot2009}  More radically, it has been shown that fundamental materials properties such as electron localisation \cite{Okamoto2007, Rini2007, Tobey2008} and superconductivity \cite{Fausti2011, Mitrano2016} can be `switched on' transiently under intense illumination.    Recently, the possibilities of manipulating materials with light have been greatly expanded by the availability of high-power sources in the 0.1--30 THz frequency range, both from table-top and free-electron lasers.\cite{Sell2008, Hoffmann2011, Liu2017}  Such developments have led to the demonstration of mode-selective optical control, whereby pumping a single infrared-active (IR) phonon mode results in a structural/electronic distortion of a second, anharmonically coupled Raman mode --- a mechanism that was termed `nonlinear phononics'.\cite{Forst2011}   Crucially, the Raman distortion is partially rectified, meaning that it oscillates around a different equilibrium position than in the absence of illumination.  

Some important light-induced phenomena such as light-induced superconductivity in C$_{60}$ (Ref. \onlinecite{Mitrano2016}) seem to depend critically on breaking molecular symmetries. \cite{Mitrano2016}  However, for mode-selective experiments, the symmetries and degeneracies of the IR and Raman modes and their implication on the transient materials properties have not thus far been discussed in a general sense, although some symmetry analysis for specific cases was performed in both experimental and theoretical papers.  For example, in the very first non-linear phononics experiment,\cite{Forst2011} performed on the rombohedrally distorted perovskite La$_{0.7}$Sr$_{0.3}$MnO$_3$, the pumped IR phonons had $E_u$ symmetry, while the non-lineary coupled Raman phonons had symmetry $A_g$ or $E_g$, depending on the pump frequency, but the `rectified' parameter (reflectivity) was probed as a scalar.  Later, the same authors performed a pump-probe X-ray Bragg diffraction experiment on La$_{0.7}$Sr$_{0.3}$MnO$_3$, in which they demonstrated the transient displacive reaction of the $E_g$ Raman mode, driven to large amplitudes by mid-infrared radiation.\cite{Forst2013}  Likewise, in the case of light-induced superconductivity in the stripe-ordered cuprate La$_{1.675}$Eu$_{0.2}$Sr$_{0.125}$CuO$_4$, a degenerate in-plane phonon was pumped in the mid-IR with 15-$\mu$m wavelength pulses, but no explicit probe of crystal symmetry breaking was sought.   In a very recent experiment, a transient enhancement of the magnetisation was achieved in the ErFeO$_3$ orthoferrite by pumping of two nearly-degenerate phonons,\cite{Nova2016} resulting in a beat of $\sim$ 1 ps.  This effect was consistent with an imaginary quadratic magneto-electric susceptibility, implying a circularly polarised phononic field at some positions within the sample.  However, neither crystallographic nor magnetic symmetry was broken in a time-average sense.  The specific case of magnetic perovskites was analysed in detail based on symmetry and density-functional theory (DFT) by Juraschek \textit{et al.} \cite{Juraschek2017}, who concluded that transient symmetry breaking by pumping nearly degenerate phonons in orthoferrites can be achieved.  Finally, in a very recent paper, Subedi discusses the possibility of inducing ferroelectricity by non-linear phononics in the mid-infrared by exploiting a quartic term in the phonon coupling.\cite{Subedi2017}
On the subject of light-induced magnetism, two very recent manuscripts by Fechner \textit{et al.}\cite{Fechner2017} and by Gu and Rondinelli \cite{Gu2017} discuss the possibility of transiently inducing ferromagnetism in Cr$_2$O$_3$ and LaTiO$_3$/YTiO$_3$, respectively, by tuning the super-exchange interactions via non-linear phononics and by direct modification of the electronic structure. \cite{Gu2017a}  These approaches are related to mine, but do not explicitly involve breaking the crystal symmetry, and as such do not seem to enable the \emph{direction} of the ferromagnetic magnetic moment to be controlled in each time-reversal antiferromagnetic domain by rotation of the pump polarisation.

Here, I discuss the possibility of achieving a transient breaking of the crystallographic or magnetic symmetry by non-linear phononics, focussing particularly on the case of \emph{trilinearly coupled, exactly degenerate IR phonons}.  Further, in the spirit of Pierre Curie's celebrated principle, ``c'est la dissym\'etrie qui cr\'ee le ph\'enom\`ene'' (it is dissymmetry that creates the phenomenon) I discuss the transient emergent \emph{ferroic} phenomena arising from breaking symmetry with light, and I propose different classes of candidate materials in which such phenomena could be investigated.  Arguably, exactly degenerate IR phonons are just a special case of the generic three-phonon coupling, and the symmetry analysis is not substantially different in the two cases.  However, I believe there are strong conceptual and experimental reasons to focus on the exactly degenerate case.  First, with degenerate IR phonons, ferroic properties can be exactly `rectified', so that they decay with the square of the damping envelope of the IR phonon after pumping.  By contrast, non-degenerate phonons will `beat' with the difference between the IR frequencies, and the associated ferroic property is likely to change in direction and/or sign through the damping envelope.  Perhaps even more importantly, the relative \emph{phase} of two non-degenerate IR phonons cannot be controlled with precision, either during the drive phase (at least one of them is necessarily driven off resonance) or during the propagation, since they travel at different speeds (this is clearly discussed, for instance, in Ref. \onlinecite{Nova2016}).   For all these reasons, I believe that photo-ferroic experiments should focus, at least initially, on systems with degenerate phonons.  I will focus primarily on \emph{mode-selective} experiments to be performed at an infra-red phonon resonance.  However, my derivation is based purely on symmetry considerations and is therefore completely general.  In the final part of this paper, I apply the same symmetry analysis to previous femto-magnetism experiments, performed  in the visible and near-infrared ranges, but sharing similar symmetry characteristics, and discuss the implications for the interpretation of these experiments.

\section{Raman rectification with degenerate phonons }

\subsection{Driving a degenerate IR mode}
As a starting point, I discuss the extension to the case of degenerate phonons of the analysis of the dynamical phonon behaviour in the presence of a \emph{cubic} coupling term between infra-red (IR) and Raman (R) phonons\footnote{ Even in the absence of inversion symmetry, I will denote as IR the phonon that is being pumped, and with Raman the one that is partially rectified.} (see ref \onlinecite{Subedi2014}, especially the supplementary information).

First of all, one can observe that the equation of motion for the IR mode (ignoring the coupling term with the Raman mode) are decoupled for a degenerate mode:

\begin{equation}
\ddot{Q}_{IR}^i+\Omega_{IR}^2Q_{IR}^i=F^i\Phi(t) \sin(\Omega t+\phi^i)
\end{equation}

where $Q_{IR}^i$ is the amplitude of the $i^{th}$ component of the degenerate mode and $F^i\Phi(t) \sin(\Omega t+\phi^i)$ is the corresponding driving term (having the dimension of a force divided by a mass), in which a phase has been explicitly included.  Whence, following the derivation of Ref. \onlinecite{Subedi2014} in the impulse approximation and assuming a single envelope function $\Phi(t)=e^{-t^2/2 \sigma^2}$ for all driving terms,

\begin{eqnarray}
{Q}_{IR}^i(t)&=&Q_{IR, max}^i \cos(\Omega t+\phi^i)\nonumber\\
Q_{IR, max}^i &=&-\sqrt{2 \pi}F^i\Omega_{IR}\sigma^3
\end{eqnarray}

Here the index $i$ runs over the degeneracy of the IR phonon (1,2,or 3 for $A/B$, $E$, and $T$ phonons, respectively).  Moreover, it is always possible to choose orthogonal effective dipole moments for the degenerate basis set, so that for an appropriate choice of the coordinates

\begin{equation}
F^i=\mathcal{Z}^*E^i
\end{equation}

where $E^i$ is the amplitude of the electric field of the light in the direction corresponding to the polarisation of mode $i$ and $\mathcal{Z}^*$ is the Born effective charge of the mode divided by its effective mass. 

Although this has no bearing on subsequent developments in this paper, it is important to emphasise that real experimental conditions are generally rather far from the impulse approximation employed by Ref. \onlinecite{Subedi2014}, and are in many cased better approximated by neglecting the damping, so that $Q_{IR, max}^i$ scales linearly with the pulse duration.  For a discussion of these cases, we defer to the treatment in Ref. \onlinecite{Forst2011}.  

\subsection{Lagrangian for the coupled IR/R modes}

In the absence of damping and forcing, we can write the Lagrangian of the coupled motion for the IR and R modes (cubic term only) as:

\begin{eqnarray}
\label{eq: Lagrangian}
\mathcal{L}&=&\frac{1}{2}\sum_i {\dot{Q}_{IR}^i}{}^2+\frac{1}{2}\sum_k {\dot{Q}_{R}^k}{}^2-\frac{\Omega_{IR}^2}{2}\sum_i {{Q}_{IR}^i}{}^2\nonumber\\
&&-\frac{\Omega_{R}^2}{2}\sum_k {{Q}_{R}^k}{}^2+G_{kij}Q^k_{R}Q^i_{IR}Q_{IR}^j
\end{eqnarray}

where the indices $i$ and $j$ run over the degeneracy of the IR mode and the index $k$ runs over the degeneracy of the Raman mode, while $G_{kij}$ is a tensor in mode space.  The Euler-Lagrange equations for the Raman coordinates then yield:

\begin{equation}
\ddot{Q}_R^k+\Omega_R^2Q_R^k-\sum_{i,j}G_{kij}Q_{IR}^iQ_{IR}^j=0
\end{equation}

Exploiting the fact that

\begin{eqnarray}
 &&\cos (\Omega_{IR}t+\phi^i)  \cos (\Omega_{IR}t+\phi^j) \nonumber\\
 &=&\frac{1}{2}\left(\cos(2 \Omega_{IR}t+\phi^i+\phi^j) +\cos \Delta \phi_{ij} \right)
\end{eqnarray}

we obtain the average (rectified) displacement of the Raman mode:

\begin{eqnarray}
\label{eq: R_modes_driven}
Q_R^k&=&\frac{1}{2 \Omega_R^2}\sum_{i,j}G_{kij}Q_{IR, max}^iQ_{IR, max}^j\cos \Delta \phi_{ij} \nonumber\\
&=&  \pi {\mathcal{Z}^*}^2 \sigma^6 \left(\frac{\Omega_{IR}^2}{\Omega_R^2}\right)\sum_{i,j}G_{kij}E^iE^j\cos \Delta \phi_{ij}
\end{eqnarray}

The static component of the Raman mode will decay with the \emph{square} of the damping envelope of the IR mode.\cite{Subedi2014}. The actual lifetime of the IR phonon is difficult to estimate with any accuracy, but with 5\% damping, the Raman envelope would be $\sim$0.5 ps for a 20 THz IR pump.  The dynamic part of the Raman mode will typically be excited very far from resonance, and will impose an oscillation around the average (static) value, as shown, for instance, in Ref. \onlinecite{Subedi2014}, Supplemental Fig. 2.

\subsection{Symmetry considerations}

The Lagrangian in Eq. \ref{eq: Lagrangian} must necessarily be totally symmetric by all elements of the point group (crystal class) of the material in question.  One consequence of this is that the tensor $G_{kij}$ must be internally symmetric in the last two indices.\footnote{Additional internal symmetries arise in the absence of dispersion --- the so-called Kleinmann's rule, but this is not relevant for subsequent discussions}  Moreover, denoting with $\Gamma_{IR}$ and  $\Gamma_{R}$ the irreducible representations (\textit{irreps}) for the IR and Raman modes, respectively, in order for any of the tensor elements to be non-zero, it must be:

\begin{equation}
\label{eq:  irrep_tensor_product}
[\Gamma_{IR} \times \Gamma_{IR}] \times \Gamma_R \sqsupset A_{1g}
\end{equation}

where $\times$ denotes the  tensor product of representations.  Equation \ref{eq:  irrep_tensor_product} reads: the symmetric part of the \emph{square} of the IR \textit{irrep} times the Raman \textit{irrep} must contain the totally symmetric \textit{irrep}.  
One trivial way to satisfy Eq. \ref{eq:  irrep_tensor_product} is for $\Gamma_R \equiv A_{1g}$, because $[\Gamma_{IR} \times \Gamma_{IR}] $ \emph{always} contains $A_{1g}$ (in fact, it contains it exactly once).  In particular, if $\Gamma_{IR}$ is non-degenerate, its square \emph{is} $A_{1g}$, so the only possible cubic coupling term for a non-degenerate IR phonon is with a totally symmetric Raman phonon.  This is the form of coupling that has been predominantly discussed thus far in the context of non-linear phononics.  In this case, the rectified Raman mode does not break any of the symmetries of the crystal, but is nonetheless capable of producing very interesting phenomenology, which may be completely inaccessible by other means.\cite{Forst2011}  However, if two IR modes are pumped simultaneously\cite{Juraschek2017} or if the IR phonon is \emph{degenerate}, the situation is significantly altered, because coupling with other Raman \textit{irreps} becomes possible.  For example in the case of a doubly-degenerate $E_u$ IR mode, $[\Gamma_{IR} \times \Gamma_{IR}]$ is a three-dimensional \emph{reducible} representation.  In addition to $A_{1g}$, it contains either a doubly-degenerate $E_g$ Raman mode or two non-degenerate Raman modes (either $A_{ng}$ with $n \ne 1$ or $B_{ng}$).  Next, I will present examples of these two cases.

\subsection{First example: point group $\bar{3}m$ ($D_{3d}$)}
\label{sec: 3barm}

The IR \textit{irrep} of interest is $E_u$.  Since $[E_u \times E_u] = A_{1g}+E_g$, a totally symmetric combination can be constructed with $\Gamma_R \equiv E_g$.  Denoting $Q_{IR}^x$,$Q_{IR}^y$ the two components of the IR mode (with polarisations along the two-fold axis and the mirror plane, respectively) and $Q_R^a$,$Q_R^b$ the two components of the Raman mode transforming with matched matrices\footnote{It is important to emphasise that $Q_R^a$ and $Q_R^b$ are \emph{not} interchangeable.  In fact, in order for the expression in Eq. \ref{eq:  totally_symmetric_3bar} to be totally symmetric, $Q_R^a$ must transform with the same matrices as $\left({Q_{IR}^y}^2-{Q_{IR}^x}^2\right)$, so, for example, it must be symmetric by a mirror plane parallel to $y$.  Likewise, $Q_R^b$ must transform with the same matrices as $2Q_{IR}^xQ_{IR}^y$, so,  for example, it must be \emph{anti}symmetric by a same mirror plane parallel to $y$.  This immediately shows that $Q_R^a$ and $Q_R^b$ have different transformation rules.}, the totally symmetric coupling term is of the form:

\begin{equation}
\label{eq:  totally_symmetric_3bar}
\left({Q_{IR}^y}^2-{Q_{IR}^x}^2\right)Q_R^a+2Q_{IR}^xQ_{IR}^yQ_R^b
\end{equation}

whence, using Eq. \ref{eq: R_modes_driven}, and noting that $\Delta \phi_{ij}=0$ for $i=j$ we find the rectified Raman modes:

\begin{eqnarray}
\label{eq: -3m_raman}
Q_R^a&=& \pi {\mathcal{Z}^*}^2 \sigma^6 \left(\frac{\Omega_{IR}^2}{\Omega_R^2}\right)(E_y^2-E_x^2)\nonumber\\
Q_R^b&=& \pi {\mathcal{Z}^*}^2 \sigma^6 \left(\frac{\Omega_{IR}^2}{\Omega_R^2}\right)(2E_xE_y \cos \Delta \phi)
\end{eqnarray}

where $\Delta \phi$ is the phase difference between $x$ and $y$ components.  From this, one can see that if we pump only along one direction ($x$ or $y$), only the $Q_R^a$ mode is rectified, but we can change its sign depending on which direction we pump ($x$ or $y$).  If we pump the two modes simultaneously with the \emph{same} amplitude the $Q_R^a$ mode is not rectified regardless of the phase $\Delta \phi$, while the $Q_R^b$ mode is rectified if $\Delta \phi \ne \pi/2+n \pi$ (i.e., if the light is not circularly polarised), and its sign can be changed by the choice of the phase ($\pm$ 45 $^{\circ}$ give opposite signs).

Both $Q_R^a$ and $Q_R^b$ break the three-fold symmetry, implying that \emph{the crystal symmetry is transiently lowered to monoclinic} ($2/m$ for $Q_R^a$, $\bar{1}$ for $Q_R^b$).    Regarding the change in sign of the rectified Raman coordinates by different choices of the IR polarisation, it is worth noting that there is no symmetry operator of the original point group that transforms $Q_R^a \rightarrow -Q_R^a$ (whereas there is one that transforms $Q_R^b \rightarrow -Q_R^b$).  Therefore, pumping along the $x$ and $y$ directions, which are not related by symmetry, will result in a physically different effect.

\subsection{Second example: point group $4/mmm$ ($D_{4h}$)}
\label{sec: 4overrmmm}

Once again, the IR \textit{irrep} of interest is $E_u$, but the decomposition is different.  Since $[E_u \times E_u] = A_{1g}+B_{1g} + B_{2g}$, we have a choice of two one-dimensional Raman modes, both antisymmetric by 90$^{\circ}$ rotation, which are clearly non-degenerate.    $B_{1g}$ is \emph{even} by the vertical and horizontal mirrors and \emph{odd} by the diagonal mirror, while the opposite is true for $B_{2g}$.  Using the same conventions as before for the IR modes and using $Q_R^a$ for  $B_{1g}$ and $Q_R^b$ for $B_{2g}$, the correct invariant has the same form as before: $({Q_{IR}^y}^2-{Q_{IR}^x}^2)Q_R^a+2Q_{IR}^xQ_{IR}^yQ_R^b$, and the equations for the rectified Raman modes are the same as in Eq. \ref{eq: -3m_raman}.  We can now see that if we pump only along $x$ or $y$ we rectify the $B_{1g}$ mode, while if we pump both directions simultaneously with the same amplitude we rectify only $B_{2g}$.  In both cases, the low-symmetry point group is $2/mmm$.  Here, none of the sign changes represents a physically different situation, because there is always a symmetry operator of the original point group that inverts the $B$ modes (in fact, the four-fold axis inverts both). 

\subsection{Parity and time reversal}

As we have seen, it is possible to break some rotational symmetries transiently by exploiting the cubic coupling between degenerate IR and Raman modes.  However, it should be immediately clear that if parity (i.e., inversion, operator symbol $\bar{1}$) is an element of the crystal class, this mechanism cannot break it.  In fact, all \textit{irreps} in the decomposition of $[E_u \times E_u]$ are \textit{gerade} (parity-even), so the Raman mode must also necessarily be \textit{gerade}.     Likewise, if time reversal (operator symbol $1'$) is an element of the magnetic point group of the crystal, this symmetry cannot be broken by the mechanism we described.  It is therefore not possible to generate magnetisation in a paramagnetic crystal, as it is the case, for instance, for the inverse Faraday effect. \cite{Shen2003}  In spite of this, as we shall see in the remainder, it is entirely possible to generate both electrical polarisation and magnetisation transiently in certain materials by non-linear degenerate phononics.

\section{Light-induced ferroic properties}

\subsection{Tensors for light-induced ferroicity}

In this section, I will discuss the \emph{transient emergence} of ferroic properties such as ferroelectricity and ferromagnetism due to light-induced symmetry breaking.  The main focus will be on properties that are otherwise absent, but I will also discuss light-induced magnetisation rotation in the important case of yttrium iron garnet (YIG).  Many of the magnetic and non-magnetic tensor calculations in this section were performed with the help of the tools in the Bilbao Crystallographic Server \cite{Perez-Mato2015, Aroyo2006}.

The simplest case is one in which the ferroic quantities (polarisation, magnetisation) are \emph{proportional} to the amplitude of the Raman modes, i.e:

\begin{eqnarray}
\label{eq: ferroic}
P_n&=&\sum_k \chi^{el}_{nk} Q_R^k\nonumber\\
M_n&=&\sum_k \chi^{mag}_{nk} Q_R^k
\end{eqnarray}

where $n$ labels the three components of the polarisation/magnetisation.  By combining Eq. \ref{eq: ferroic} with Eq. \ref{eq: R_modes_driven} we obtain:

\begin{eqnarray}
\label{eq: ferroic2}
P_n&=&\pi {\mathcal{Z}^*}^2 \sigma^6 \left(\frac{\Omega_{IR}^2}{\Omega_R^2}\right) \sum_{ij} T^{el}_{nij}E^iE^j\cos \Delta \phi_{ij} \nonumber\\
M_n&=&\pi {\mathcal{Z}^*}^2 \sigma^6 \left(\frac{\Omega_{IR}^2}{\Omega_R^2}\right) \sum_{ij} T^{mag}_{nij}E^iE^j\cos \Delta \phi_{ij}
\end{eqnarray}

where

\begin{eqnarray}
\label{eq: full_tensors}
T^{el}_{nij}&=& \sum_k \chi^{el}_{nk} G_{kij} \nonumber\\
T^{mag}_{nij}&=& \sum_k \chi^{mag}_{nk} G_{kij} 
\end{eqnarray}

A key observation here is that the $T$ tensors in Eq. \ref{eq: full_tensors} are simply restrictions to the plane of polarisation of the degenerate phonons of more general materials tensors, which are symmetric in the last two indices $i$ and $j$.  Moreover, $T^{el}_{nij}$ is time-reversal-even and parity odd, whereas $T^{mag}_{nij}$ is time-reversal-odd and parity-even --- in other words, they transform like the \emph{piezoeletric tensor} and the \emph{piezo-magnetic tensor}, respectively.   This immediately suggests a useful search strategy for materials that might display light-induced ferroicity:  materials that might display light-induced \emph{electrical polarisation} must be \emph{piezoelectric}, while materials that might display light-induced \emph{magnetisation} must be \emph{piezomagnetic}.  Moreover, $T^{mag}_{nij}$ also transforms like the $HEE$ quadratic magneto-electric tensor, which is the \emph{static} version of the dynamic magnetisation effect we just described (note that the related $EHH$ effect is not relevant here).  Therefore, materials displaying the  $HEE$ quadratic magneto-electric effect might be considered prime candidates for light-induced magnetisation.

In spite of the similarities with well-known static effects, it is important to emphasise three unique features of dynamical experiments with light;
\begin{enumerate}
\item The most obvious advantage of dynamical experiments with light is \emph{the time structure of the pump carrier envelope}, which should enable  transient ferroic effects of $\lesssim$ 1 ps to be produced.
\item Whereas static effects generally involve all phonon modes with a weight that cannot be controlled, with non-linear phononics one can select different IR phonon resonances by changing the wavelength of the pump.  Although the form of the tensors for IR and Raman phonons of a given symmetry is always the same, the magnitude and sign of the effect could be very different for different phonons.
\item In static experiments, changing the direction of the strain or the applied electric field usually requires remounting the sample of even preparing crystals with different cuts.  By contrast, in dynamic light experiments, different electric field orientations can simply be achieved by rotating the polarisation of the pump.  As we have seen in sections \ref{sec: 3barm} and \ref{sec: 4overrmmm}, rotating the polarisation of the light enables one to reverse the effect or probe different physical situations \textit{in situ}.
\end{enumerate}

\subsection{Light-induced ferroelectricity}

The possibility of inducing ferroelectricity in a non-polar material transiently at time scales of the order of a ps or less is intriguing.  If the effect is sufficiently large, one might envisage applying it, for example, in the dynamical equivalent of the Ferroelectric Field-Effect Transistor\cite{Meena2014} --- a device that could be opened for an extremely short time (as short as a single cycle of the pump) and would immediately return to the `off' state as for photoconductive Auston switches but with much lower pump photon energy.

Exactly as in the case of piezoelectricity, light-induced ferroelectricity is only permitted in 20 of the 21 non-centrosymmetric crystal classes.  In the 10 pyroelectric (polar) classes, polarisation is already present, which makes it more difficult to detect light-induced ferroelectricity, for example, through second-harmonic generation (SHG).  Of the remaining 10 crystal classes, $222$ does not have any degenerate modes, while $422$ and $622$ have zero tensor components in the plane of the degenerate modes, so these classes can be excluded.  Here, we focus on uniaxial point groups with E modes in the $ab$ plane. For these groups, the relevant tensor elements are $T^{el}_{.1}$ ($x^2$), $T^{el}_{.2}$ ($y^2$) and $T^{el}_{.6}$ ($2xy$), where we have employed the usual notation for tensors that are symmetric in the last two indices:  $T^{el}_{11} \equiv T^{el}_{111}$, $T^{el}_{14} \equiv 2T^{el}_{123}$, etc.  Of particular interest are crystal classes $\bar{4}$ ($S_4$) and $\bar{4}2m$ ($D_{2d}$), since they have elements of the type $T^{el}_{36}$ (both classes) and $T^{el}_{31}, T^{el}_{32}$ (only $\bar{4}$), implying that the light-induced polarisation will develop along the high-symmetry $c$ axis.

In addition to having the appropriate symmetry, candidate materials should also have other characteristics:  they should be good insulators with large gaps, and should possess IR degenerate modes that can be easily pumped.  Although recent developments in THz generation have achieved significant intensities with narrow carrier envelopes in the THz gap, much higher THz generation efficiencies can be achieved above 20 THz, considerably simplifying this type of experiments.  It is therefore desirable to explore in the first instance materials that are entirely composed of light elements, since they generally possess high IR phonon frequencies. 

\subsubsection{Possible light-induced ferroelectricity in BPO$_4$}

BPO$_4$ crystallises in space group $I\bar{4}$ (82), point group $\bar{4}$  ($S_4$) (Ref. \onlinecite{Schmidt2004}).  Its structure is similar to that of $\alpha$-crystobalite, and consists of corner-sharing tetrahedra of BO$_4$ and PO$_4$, all with point-group symmetry $\bar{4}$ and therefore non-polar. BPO$_4$  has a large band gap of the order of 10.5 eV (Ref. \onlinecite{Li2016}).  IR \cite{Dultz1975, Adamczyk2000} and Raman \cite{Osaka1984, Adamczyk2000} studies indicate the presence of several IR- and Raman-active modes above 1000 cm$^{-1}$ ($\sim$30 THz).  Although the symmetry assignment of the early experimental data was somewhat ambiguous, theory predicted four modes of symmetries $E$, $B_1$, $E$, and $B_2$ with wavenumbers in the range 1087--1195 cm$^{-1}$ (Ref. \onlinecite{Osaka1984}).  In this frequency range, it is possible to obtain peak electric field amplitudes in excess of 40 MV/cm.  Applying a compressive strain or an oscillating electric field along the $a$ axis will reduce the symmetry of the tetrahedra to $2$ (polar).  The polarisations of the distorted tetrahedra points up/down for BO$_4$ and PO$_4$--tetrahedra, respectively, but the two dipole moments will not compensate, giving rise to a net polarisation, which is expected to be reversed if the strain is along the $b$ axis.  Strain along the $[110]$ or $[\bar{1}10]$ direction will produce a qualitatively similar effect but with a different magnitude.  This is confirmed by  symmetry analysis:  since $\bar{4}$ is an Abelian group, all its \textit{irreps} are one-dimensional, but two of them ($^1$E and $^2$E, complex conjugate of each other) combine into a physically irreducible two-dimensional \textit{irrep} $E_{phys}$.   Since $[E_{phys} \times E_{phys}] = A+2B$, we can write two independent invariants containing the square of the electric field and the 'Raman' mode $Q_R$ transforming like $z$.  This mode, which reduces the symmetry from $\bar{4}$ to $2$, carries a finite polarisation, and is here allowed because the material is non-centrosymmetric in the first place. The coupling term will be of the form $c_1({Q_{IR}^y}^2-{Q_{IR}^x}^2)Q_R+c_2(2{Q_{IR}^x}{Q_{IR}^y})Q_R$, with $c_1$ and $c_2$ being materials-specific coefficients.  Both the static polarisation induced by application of an in-plane electric field and the rectified, light-induced polarisation will then be directly proportional to the amplitude of $Q_R$.  For light polarised in the $ab$ plane and having a generic elliptical polarisation with $E_x=E \cos \theta$, $E_y=E \sin \theta e^{i \Delta \phi}$ we obtain

\begin{equation}
\label{eq: BPO4_Pol}
P_z \propto \frac{1}{2}E^2(c_1 \cos 2 \theta+c_2 \sin 2 \theta \cos \Delta \phi)
\end{equation}

It is useful to compare the polarisation in Eq. \ref{eq: BPO4_Pol} with the one obtained from the piezoelectric tensor in point group $\bar{4}$:

\begin{equation}
\label{eq: BPO4 Tensor}
D_{\bar{4}}=\left(\begin{array}{cccccc} 
0&0&0&d_{14}&d_{15}&0\\
0&0&0&-d_{15}&d_{14}&0\\
d_{31}&-d_{31}&0&0&0&d_{36}\end{array}\right)
\end{equation}

As previously stated, the mode-selective photo-induced polarisation can be obtained from the \emph{restriction} to the plane of polarisation of the $E$ mode of a tensor with the same form as that in Eq. \ref{eq: BPO4 Tensor}, in this case with the correspondences $c_1=d_{31}$ and $2c_2=d_{36}$.  One might therefore wonder whether the remaining elements of the tensor would give rise to mode-selective, light-induced polarisation.  The answer to this question is that such an effect would require pumping \emph{simultaneously and coherently} modes with $ab$ and $z$ polarisation, which clearly cannot be degenerate in this symmetry.  Nevertheless, it might be possible in some cases to activate both modes with a broadband pulse, if the frequencies are not too different --- an approach that was followed in Ref. \onlinecite{Nova2016}.

In addition to piezoelectricity, another static effect that is analogous to light-induced polarisation and is described by the same tensor form as Eq. \ref{eq: BPO4 Tensor} is \emph{quadratic dielectricity}, meaning the development of an electrical polarisation that is quadratic in the applied electric field.  Calculations performed in Ref. \onlinecite{Li2016} on BPO$_4$ in fact predict a $z$-axis polarisation that is \emph{quadratic} in the electric field applied in the $xy$ plane.  The calculation in Ref. \onlinecite{Li2016} was only performed in the $[110]$ and $\bar{1}10]$ directions, so only the static $c_2$ coefficient was determined, yielding $c_2 \sim 1.56 \times 10^{-22}$ in SI units.  According to these calculations, for a \emph{static} electric field amplitude of 40 MV/cm, one would obtain a polarisation of $\sim 2500 \mu C/m^2$.  This is to be compared with BaTiO$_3$ ($\sim 300,000 \mu C/m^2$) and TbMnO$_3$ ($\sim 800 \mu C/m^2$) --- so it is not large compared to ordinary ferroelectrics but is would be easily measurable in a static setting.

As previously explained, it is very difficult to predict the outcome of an ultra-fast experiment by extrapolating from static measurements.  A better assessment of the response of each IR mode could be obtained from first-principles calculations, similar to Ref. \onlinecite{Juraschek2017}.  Nevertheless, these numbers suggest that light-induced polarisation may be observable and may even be large.  In the first instance, the best approach seems to be to use SHG as a probe of the symmetry reduction.  The SHG tensor in the presence and absence of illumination can be expressed as:

\begin{equation}
\chi^{(2)}_{on}=\chi^{(2)}_{off}+\delta\chi^{(2)}
\end{equation}

where

\begin{equation}
\chi^{(2)}_{off}=\left(\begin{array}{cccccc} 
0&0&0&e_{14}&e_{15}&0\\
0&0&0&-e_{15}&e_{14}&0\\
e_{31}&-e_{31}&0&0&0&e_{36}\end{array}\right)
\end{equation}

and

\begin{equation}
\delta\chi^{(2)}=\left(\begin{array}{cccccc} 
0&0&0&\delta_3&\delta_4&0\\
0&0&0&\delta_4&-\delta_3&0\\
\delta_1&\delta_1&\delta_2&0&0&0\end{array}\right)
\end{equation}

One can see that the only new element of the SHG tensor is $\chi^{(2)}_{33}$, which was strictly zero in the ``off'' state.  This element allows light at $2 \omega$ with $z$ polarisation to be produced by light at $\omega$ also having $z$ polarisation. 

\subsubsection{$\gamma$-LIBO$_2$ --- another potential candidate for light-induced ferroelectricity}
$\gamma$-LIBO$_2$ is a high-pressure polymorph of LIBO$_2$ (lithium metaborate), which can be grown as large single crystals in hydrothermal conditions.\cite{McMillen2008}  $\gamma$-LIBO$_2$ crystallises in space group $I\bar{4}2d$ (122), i.e., point group $\bar{4}2m$ ($D_{2d}$), which is the second candidate point group I previously discussed.  To my knowledge, the IR absorption spectra of $\gamma$-LiBO$_2$ has not been measured, but the IR spectra of the ordinary (monoclinic) form contain several bands in the region 1000--1400 cm$^{-1}$.  In point group $\bar{4}2m$, the only element of the piezoelectric tensor that produces a polarisation along the $c$ axis is $d_{36}$, meaning that the strain (or pump polarisation for light-induced FE) must be along the $[110]$ or $[\bar{1}10]$ directions.  This is confirmed by the \textit{irrep} decomposition:  there is a single doubly-degenerate $E$ \textit{irrep}, and the symmetric part of its square contains two $B$ modes:  $[E \times E] = A+B_1+B_2$.  The coupling term is of the form $c_1({Q_{IR}^y}^2-{Q_{IR}^x}^2)Q_R^a+c_2(2{Q_{IR}^x}{Q_{IR}^y})Q_R^b$, where $Q_R^a$ and $Q_R^b$ transform like $B_1$ and $B_2$, respectively, and $c_1$ and $c_2$ are materials-specific coefficients.  The $B_2$ modes reduces the symmetry to the ferroelectric point group $mm2$, producing electrical polarisation along the $c$-axis. By contrast, the $B_1$ mode reduces the point group symmetry to $222$, which is \emph{not} ferroelectric.  Consequently, light-induced polarisation will only develop if both $E_x$ and $E_y$ are non-zero and are not out of phase.

The piezoelectric tensor in this crystal class has the form:

\begin{equation}
\label{eq: LiBO4 Tensor}
D_{\bar{4}2m}=\left(\begin{array}{cccccc} 
0&0&0&d_{14}&0&0\\
0&0&0&0&d_{14}&0\\
0&0&0&0&0&d_{36}\end{array}\right)
\end{equation}

which, once restricted to the $ab$ plane, yields exactly the same result for the electrical polarisation.

\subsection{Light-induced ferromagnetism}

The idea of light-induced ferromagnetism is very similar to that of light-induced polarisation: to produce a magnetisation using the rectified component of Raman mode, which would break symmetries other than time reversal ($1'$).  The analogues of piezoelectricity and quadratic dielectricity in this context are \emph{piezomagnetism} and the \emph{static quadratic magneto-electric} ($HEE-ME$) effect, so one should look in the first instance for materials that are piezomagnetic or $HEE-ME$ but not already ferromagnetic.  Note that all such piezomagnetic materials are also antiferromagnetic, because the piezomagnetic tensor must be time-reversal odd.  Moreover, any static or dynamic effect of this kind changes sign for time-reversal antiferromagnetic domains, so one always relies on natural or induced domain imbalance (see discussion further below).

If these effects turned out to be large, one could envisage possible applications in the field of ultra-fast magnetic switches and photo-antennae for spin wave generation.  The additional materials-specific properties required for pump-probe experiments are very similar to those for light-induced polarisation, but are mode difficult to fulfil: since a magnetic element is required, these materials cannot be made entirely of very light elements.  We may therefore expect to have to pump in the few THz range.  However, these materials should be rather easy to probe, because Faraday rotation is a very sensitive probe of magnetism.  If previous static measurements of the $HEE$ effect exist, they can be employed to provide a rough estimate of the magnitude of the light-induced effect, as previously discussed for BPO$_4$.

\subsubsection{MnF$_2$ and CoF$_2$ --- two known piezomagnetic materials}

The problem of piezomagnetism in MnF$_2$ and CoF$_2$ was already investigated in the late 1950s, both theoretically \cite{Dzialoshinskii1957} and experimentally. \cite{Borovik-Romanov1960}  These materials are antiferromagnets with N\`eel temperatures of $\sim$ 66.5 K and $\sim$ 37.7 K, respectively.  They both crystallise in the rutile structure, with magnetic space group $P4'_2/mnm'$ (magnetic point group $4'/mmm'$), which does not support ferromagnetism.  The magnetic moments are along the $c$ axis, and cancel exactly in the two octahedra at the origin and at the centre of the cell, respectively, which are rotated by 90$^{\circ}$.   

The piezomagnetic tensor has only two non-zero independent components:\cite{Perez-Mato2015}  $\Lambda_{14}=\Lambda_{25}$ and $\Lambda_{36}$.  The first two elements describe the canting of the moment away for the $z$ direction upon application of stress with a $\sigma_{yz}$ or $\sigma_{xz}$ component, giving rise to a net moment in the $x$ or $y$ directions, respectively, and are not relevant here because the corresponding IR phonons would not be degenerate.  By contrast, $\Lambda_{36}$ describes the application of stress in the $ab$ plane, and would correspond to pumping two in-plane degenerate phonons. Application of stress along $a$ or $b$ does not result in a ferromagnetic (FM) moment, because the symmetry is reduced to point group $222$ (non-FM).  However, application of stress along the $[110]$ direction results in an induced FM moment along the $c$ axis (FM point group $m'm'm$).  The reason for this behaviour is that strain in the $[110]$ direction makes the two octahedra inequivalent (apically extended/compressed), resulting in a small change in magnetic moment via spin-orbit coupling --- an effect that should be particularly prominet in the Co compunds.  The two moments in the unit cell no longer cancel perfectly, giving rise to strain-induced \emph{collinear ferrimagnetism}.

As previously discussed, one expects light-induced ferromagnetism to be described by a tensor that has exactly the same form of the piezo-electric tensor.  Therefore, the rectified magnetisation would be maximum for polarisation in the $[110]$ direction, and would change sign by for 90$^{\circ}$ rotation of the polarisation or for reversal of the staggered magnetisation (time-reversal domains).  Appropriate $E_u$ modes in CoF$_2$ were previously measured, and are in the range 200-450 cm$^{-1}$ (6-15 THz), \cite{Balkanski1966} while slightly higher frequencies are expected for MnF$_2$ .  

Although not strictly relevant for mode-selective experiments, the recent experiment performed by Higuchi \textit{et al.} on MnF$_2$ in the visible range (Ref. \onlinecite{Higuchi2016}) is worth discussing in this context, because it demonstrates a polarisation-dependent (azimuth) effect that is entirely consistent by symmetry with the one I have just described.  Higuchi \textit{et al.} showed that it is possible to control the antiferromagnetic time-reversal domain population by selective heating of one of the domains by magnetic linear dichroism (MLD), and observe that the symmetric restriction and tensor forms for an antiferromagnetic material exhibiting MLD are the same as that of the piezomagnetic effect.  The relevant tensor elements for MLD MnF$_2$ is element $36$, implying that the MLD effect is maximum for the $[110]$ direction, and that it changes sign for 90$^{\circ}$ rotation of the polarisation, reversal of the staggered magnetisation or reversal of the applied magnetic field, as indeed obvserved experimentally.

\subsubsection{Cr$_2$O$_3$ and Fe$_2$O$_3$}

In this section, I briefly discuss the 'classic' collinear antiferromagnets Cr$_2$O$_3$ and Fe$_2$O$_3$, as a useful illustration of how subtly different symmetries can give rise to different non-linear phononic responses.  Both materials crystallise in the corundum structure (space group $R\bar{3}c$, no. 167) and, in the ground state,  have spins pointing along the trigonal $c$ axis.  Cr$_2$O$_3$ has magnetic point group symmetry $\bar{3}'m'$ at all temperatures, and is is well known to display the linear magnetoelectric effect.  All elements of the piezomagnetic $HEE-ME$ tensor in this point group are zero by symmetry, so no magnetisation can be induced in Cr$_2$O$_3$  by non-linear phononics.  By contrast,  below the Morin transition ($\sim$ 260 K), Fe$_2$O$_3$ has magnetic point group $\bar{3}m$, and is well known to be piezomagnetic \cite{Phillips1967, Andratskil1967}, with the following tensor:

\begin{equation}
\label{eq: Fe2O3 Tensor}
\Lambda=\left(\begin{array}{cccccc} 
\Lambda_{111}&-\Lambda_{111}&0&2\Lambda_{123}&0&0\\
0&0&0&0&-2\Lambda_{123}&-2\Lambda_{111}\\
0&0&0&0&0&0\end{array}\right)
\end{equation}

As previously discussed, the $T^{mag}_{nij}$ tensor has exactly the same form of the piezomagnetic tensor.

For light polarisation in the plane $ab$ plane, one obtains:

\begin{eqnarray}
M_x&\propto& +E^2 T^{mag}_{111} \cos 2 \theta\nonumber\\
M_y&\propto&- E^2 T^{mag}_{111} \sin 2 \theta\\
\nonumber
\end{eqnarray}

where the $x$ axis is the direction of the two-fold axis and $\theta$ is the in-plane angle of the polarisation with respect to the $x$ direction.  Fe$_2$O$_3$ has optical modes with in-plane polarisation in the range 380--650 nm (Ref. \onlinecite{Onari1977}) and has a gap of $\sim$ 2.1 eV (Ref. \onlinecite{Marusak1980}), indicating that it could be an excellent candidate to observe photo-magnetism by non-linear phononics.

\subsubsection{The HEE-ME effect and light-induced rotation of the magnetisation in yttrium iron garnet (YIG).}

\label{sec: YIG}

Yittrium iron garnet (YIG), with chemical formula Y$_3$Fe$_5$O$_{12}$ and space group $Ia\bar{3}d$ (230),  is a well-known room-temperature ferrimagnetic material, with a variety of existing applications including in non-linear optics. \cite{Aichele2003}  Since YIG is already ferrimagnetic, photo-ferromagnetism would not be an entirely emergent property, but I make an exception here because the interpretation of previous photo-magnetism experiments\cite{Stupakiewicz2017} could, I believe, be greatly assisted by a symmetry approach.

 YIG has been known for many years to display a strong $HEE$ quadratic magnetoelectric effect \cite{ODell1967}.  In a single ferrimagnetic domain magnetised along the $[111]$ direction, the point group symmetry is reduced from cubic (paramagnetic point group $m\bar{3}m1'$) to trigonal $\bar{3}m'$.  Its  $HEE$ tensor  $M_i=\alpha_{ijk} E_jE_k$ expressed in the trigonal reference frame is of the following form\footnote{I used the tensor calculator for magnetic point groups from the Bilbao Crystallographic Server}:

\begin{equation}
\label{eq: YIG_tensor}
\alpha=\left(\begin{array}{cccccc} 
0&0&0&0&2\alpha_{113}&-2\alpha_{222}\\
-\alpha_{222}&\alpha_{222}&0&2\alpha_{113}&0&0\\
\alpha_{311}&\alpha_{311}&\alpha_{333}&0&0&0\end{array}\right)
\end{equation}

In the cubic reference frame, taking the $[111]$ direction ($\parallel$ to the magnetisation) as the $\hat{z}$ direction, we can choose $\hat{x}$ along $[1\bar{1}0]$ (parallel to the two-fold axis) and $\hat{y}$ along $11\bar{2}$ (perpendicular to the two-fold axis).  It is easy to see that an electric field along either $x$ or $y$ gives rise to a moment along $y$, whereas an electric field at 45$^{\circ}$ between $x$ and $y$ gives rise to a moment along $x$.  This is indeed what was found in  Ref. \onlinecite{ODell1967}.  The effect was easily measured  with very modest fields ($\sim$7.5kV/cm), and should be very large in the ultra-fast regime both because of the larger $E$ fields and because of the resonant enhancement.  This effect would manifest itself in several ways:  at relatively low peak fields, one would observe a transient canting of the magnetic moments away from the cubic $[111]$ direction, followed by a Gilbert-damped spin precession, during which the Faraday rotation in the original direction would be reduced.  At very high peak fields, the effect may be so large as to enable dynamic switching of domains and a permanent reduction of the Faraday rotation.\cite{Maehrlein2017}

In this context, it is interesting to discuss the results recently obtained by Stupakiewicz \textit{et al.} on Co-doped YIG, \cite{Stupakiewicz2017}, which demonstrated ultra-fast writing and erasing of magnetic `bits'.  In that experiment, the wavelength was varied within the range 1,150 -- 1,450 nm (1.08 -- 0.86 eV), where the light resonantly excites electronic d--d transitions in Co ions, so the conditions are very far from those previously discussed for mode-selective experiments.  It is also important to emphasise that the decrease of single-ion anisotropy induced by pumping in resonance with a Co transition appears to be a crucial ingredient for the interpretation of this experiment.  However, since the effect is observed with linearly polarised light and is clearly non-linear in nature, it is reasonable to hypothesise that the symmetry of the effect might still be described by a tensor with the same form as the piezomagnetic tensor.  The remainder of this section is devoted to discussing the consequences of this hypothesis.  The analysis is performed under the assumption that the cubic symmetry is only broken in each domain by the local direction of the magnetisation along one of the cubic $\langle 111\rangle$ directions.  Strictly speaking, this assumption is incorrect, since the cubic symmetry is additionally broken by both the substrate and the miscut.  Nevertheless, I will show that this rather naive analysis provides a good explanation for the `writing' process, while other aspects of the experiment, related to the anisotropies that I have disregarded, are less well interpreted.

The starting point of the analysis is the calculation of the light induced magnetisation $\Delta \vec{m}$ for given pump polarisation ($\vec{p}=[100]$ and $[010]$ in the cubic setting) and direction of the equilibrium magnetisation $\vec{m}$ ($[1\bar{1}1]$/$[1 1 \bar{1}]$ for $M^{(L)}_+$ /$M^{(L)}_-$ domains in Ref. \onlinecite{ Stupakiewicz2017}, Fig. 2, and so on), under the symmetry assumptions discussed here above.    This requires transforming the pump direction to the trigonal setting of the tensor in Eq. \ref{eq: YIG_tensor}, applying the tensor itself and transforming back to the cubic coordinates.  This calculation is performed in detail in the Appendix, and the results are summarised here below.

\begin{enumerate}
\item  \label{pt: axes} In all cases, the transverse component of the magnetisation induces a rotation around one of the $\langle 110\rangle$ cubic directions, which is perpendicular to both  $\vec{m}$ and $\vec{p}$.  So, for example, for $\vec{m} \parallel [1\bar{1}1]$ (as for domain $M^{(L)}_+$ ) and $\vec{p}=[100]$, the axis of rotation is $\vec{a}=[011]$, while for $\vec{p}=[010]$, $\vec{a}=[10\bar{1}]$.  The direction of $\vec{a}$ defines the positive (counterclockwise) and negative (clockwise) rotation.
\item The magnitude and sign of the torque is given by a material-specific parameter $\kappa_1$ (related to the tensor elements --- see Eq. \ref{eq: kappa_def}), which is the same for both directions of $\vec{p}$, and tends to switch $\vec{m}$ towards a different body diagonal.  

\item \label{pt: kpos} For  $\kappa_1 > 0$,   $\vec{p}=[100]$ will produce $[1\bar{1}1] \rightarrow [1 1 \bar{1}]$, i.e., $M^{(L)}_+ \rightarrow M^{(L)}_-$, and likewise $M^{(S)}_- \rightarrow M^{(S)}_+$ and so on. By contrast, $\vec{p}=[010]$ will produce $[1\bar{1}1] \rightarrow [\bar{1} \bar{1} \bar{1}]$, which is disfavoured by the miscut.  

\item \label{pt: kneg} For  $\kappa_1< 0$,   $\vec{p}=[100]$ will produce $[1\bar{1}1] \rightarrow [\bar{1} \bar{1} 1]$ (a disfavoured domain))  while $\vec{p}=[010]$ will produce $[1\bar{1}1] \rightarrow [111]$, i.e., $M^{(L)}_+ \rightarrow M^{(S)}_+$.
\end{enumerate}

From this, we can see that the dynamics of the `writing process' in Stupakiewicz \textit{et al.} is well described by our $\kappa_1 > 0$ scenario.  It must be emphasised that, in the analysis I just presented, the cubic symmetry is only broken by the magnetisation, so aspects of the experiments clearly related to other forms of symmetry breaking are not well explained.   In particular, the direction of $\vec{p}$ that dynamically switches $M^{(L)}_+ \rightarrow M^{(L)}_-$ (`write') should also most efficiently switch $M^{(L)}_- \rightarrow M^{(L)}_+$(`erase') since the two domains are related by a rotation around the pump direction.  Nevertheless, the `written' domain is a metastable island in a large domain with a different magnetisation, so an 'erase' pulse in the $[010]$ direction may also restore a uniform magnetisation.

\section{Summary and conclusions}

In summary, I have presented a theory describing how crystal symmetry can be broken transiently by light in the mid-IR or THz range thorough third-order non-linear coupling, focussing on the case where two exactly degenerate IR 'pump' phonons couple with a Raman phonon --- a process that is highly relevant in the emergent field of non-linear phononics.  I have further demonstrated that, in many cases, this process can give rise to `rectified' ferroic properties such as ferroelectricity and ferromagnetism, which would emerge transiently and decay with the square of the damping envelope of the IR phonon (typically $\lesssim$ 1 ps decay time).   I discussed several classes of potential `photo-ferroelectric' and `photo-ferromagnetic' materials, possessing the appropriate symmetry requirements and other favourable properties such as IR phonons in accessible ranges, sizeable band gaps, and large cognate static effects, and could therefore be prime candidates for non-linear phononics experiments.  Finally, I discussed two previous photo-magnetism experiments on MnF$_2$  by Higuchi \textit{et al.} (Ref. \onlinecite{Higuchi2016}), and on Co-doped Yittrium Iron Garnet (YIG ---Y$_3$Fe$_5$O$_{12}$) by Stupakiewicz \textit{et al.} (Ref. \onlinecite{Stupakiewicz2017}). Although performed in the visible/near-IR range, respectively, and therefore in a completely different frequency range from that of non-linear phononics, I have shown that these experiments can be interpreted by a very similar symmetry analysis.  Should the `photo-ferroic' properties I have discussed be experimentally accessible by non-linear phononics, these and other materials could find useful applications in ultra-fast electronics and information storage.

\appendix

\section{Photo-induced magnetisation in YIG} \label{app: prove of YIG}

Although transforming the tensor in Eq. \ref{eq: YIG_tensor} to the cubic coordinates is straightforward, I believe it is more instructive to perform a step-by-step calculation.  One first defines a set of Cartesian coordinates that is appropriate for the trigonal setting.  For the magnetisation along the $[1\bar{1}1]$ direction, one such set is:

\begin{eqnarray}
\hat{x}&=&\frac{\sqrt{2}}{2} [110]\nonumber\\
\hat{y}&=&\frac{\sqrt{6}}{6} [1\bar{1}\bar{2}]\nonumber\\
\hat{z}&=&\frac{\sqrt{3}}{3} [1\bar{1}1]
\end{eqnarray}

In this reference frame, the two pump directions are:

\begin{eqnarray}
[100]&=&\frac{\sqrt{2}}{2} \hat{x}+\frac{\sqrt{6}}{6}\hat{y}+\frac{\sqrt{3}}{3} \hat{z}\nonumber\\\nonumber\\
{[1}00{]}&=&\frac{\sqrt{2}}{2} \hat{x}-\frac{\sqrt{6}}{6}\hat{y}-\frac{\sqrt{3}}{3} \hat{z}
\end{eqnarray}

Applying the tensor in Eq. \ref{eq: YIG_tensor} one obtains:

\begin{eqnarray}
\label{eq: dm100}
\Delta m_x^{[100]}&=&\sqrt{3}\kappa_1 \nonumber\\
\Delta m_y^{[100]}&=&\kappa_1 \nonumber\\
\Delta m_z^{[100]}&=&\kappa_2
\end{eqnarray}

and

\begin{eqnarray}
\label{eq: dm010}
\Delta m_x^{[010]}&=&-\sqrt{3}\kappa_1 \nonumber\\
\Delta m_y^{[010]}&=&\kappa_1 \nonumber\\
\Delta m_z^{[010]}&=&\kappa_2
\end{eqnarray}

with

\begin{eqnarray}
\label{eq: kappa_def}
\kappa_1&=&\frac{1}{3}\left( \sqrt{2}\alpha_{113}-\alpha_{222}\right)\nonumber\\
\kappa_2&=&\frac{1}{3}\left(2\alpha_{311}+\alpha_{333}\right)
\end{eqnarray}

From Eqs. \ref{eq: dm100} and \ref{eq: dm010}, one can see that for both directions of the pump there is a component of the light-induced magnetisation along the original magnetisation $\vec{m}$, which is the same for both pump directions, and the sign of which is controlled by the materials-specific parameter $\kappa_2$.  The remaining part of $\Delta \vec{m}$ is \emph{perpendicular} to $\vec{m}$ and, going back to the cubic setting, is:

\begin{eqnarray}
\Delta \vec{m}_{\perp}^{[100]}&=&\frac{\sqrt{6}}{3} \kappa_1 [21\bar{1}]\nonumber\\
\Delta \vec{m}_{\perp}^{[010]}&=&\frac{\sqrt{6}}{3} \kappa_1[\bar{1}\bar{2}\bar{1}]
\end{eqnarray}

from which one can calculate the torque vector:

\begin{eqnarray}
\vec{\tau}^{[100]}=\vec{m} \times \Delta \vec{m}_{\perp}^{[100]}=\sqrt{2}\kappa_1 [011]\nonumber\\
\vec{\tau}^{[010]}=\vec{m} \times \Delta \vec{m}_{\perp}^{[010]}=\sqrt{2}\kappa_1 [10\bar{1}]
\end{eqnarray}

For this, one can easily verify the statements in sec. \ref{sec: YIG}, points \ref{pt: axes} to \ref{pt: kneg}.

\acknowledgements{This work was funded by EPSRC grant No. EP/M020517/1, entitled ``Oxford Quantum Materials Platform Grant''.  I acknowledge invaluable discussions with Andrea Cavalleri, Tobia Nova, Michael Fechner and Michael F\"orst at the MPSD institute in Hamburg.}

\bibliography{Phononics_prb}

\end{document}